\documentstyle[12pt]{article}
\begin{document}
\def\thefootnote{\fnsymbol{footnote}}
\begin{flushright}
KANAZAWA-05-08  \\
April, 2005\\  
\end{flushright}
\vspace*{2cm}
\begin{center}
{\LARGE\bf Nonthermal production of baryon and dark matter}\\
\vspace{1 cm}
{\Large  Daijiro Suematsu}
\footnote[1]{e-mail:suematsu@hep.s.kanazawa-u.ac.jp}
\vspace {0.7cm}\\
{\it Institute for Theoretical Physics, Kanazawa University,\\
        Kanazawa 920-1192, JAPAN}
\end{center}
\vspace{2cm}
{\Large\bf Abstract}\\
We study nonthermal production of baryon and dark matter.
If we extend the MSSM by introducing some singlet
chiral superfields so as to enlarge the conserved global symmetry, 
the abundance of 
the baryon and the dark matter in the universe may be explained as 
the charge asymmetry of that symmetry. In such a case, the baryon energy 
density and the dark matter energy density in the present universe can
be correlated each other and take the similar order values naturally.   
\newpage
\setcounter{footnote}{0}
\def\thefootnote{\arabic{footnote}}
Both energy densities of baryon and dark matter in the universe
have been measured through observations of the 
CMB anisotropy \cite{cmb} and the large scale structure \cite{lss}. 
In the present astroparticle physics, it is one of crucial problems
to explain why these energy densities take the similar order values.
In the supersymmetric framework, the dark matter abundance is usually
considered to be explained through the thermal production of 
the lightest stable superparticles with the weak interaction. 
Although it is considered to be a promising possibility, 
the scenario has a large dependence on the feature 
of supersymmetry (SUSY) breakings \cite{cdm} and we also need an
independent mechanism to produce the baryon number abundance.
The SUSY breaking parameters seem to be required to be tuned 
to realize the observed abundances.

On the other hand, many years ago, an interesting idea was proposed 
in \cite{dark}, where both abundances of the baryon and the dark 
matter are related to the charge asymmetry of the same global symmetry. 
The scenario seems to be able to explain naturally why they have 
the similar order values.
Although the idea is very elegant, such a kind of realistic model
seems not to have been constructed a lot in the context of SUSY
models.\footnote{An example of this kind of works can be found in
\cite{hmw}. The production of baryons and dark matter through the
nonthermal decay process has been also discussed in the MSSM context in 
\cite{nonth}.}
In fact, since the Lagrangian of the minimal supersymmetric 
standard model (MSSM) has only baryon number $(B)$ 
and lepton number $(L)$ as its global symmetry,
it may be difficult to fulfill the required condition to realize 
that scenario.\footnote{Although one might consider to use 
$L$ or $B-L$ as the global charge discussed in the following study, 
in that case we will face the difficulty such that it cannot explain the
$B$ asymmetry and the small neutrino masses, simultaneously.} 
However, if we introduce new singlet chiral superfields in the MSSM,
the global symmetry can be extended and the idea may be applicable to
the SUSY model.

In the MSSM, both operators $\hat{H}_1\hat{H}_2$ and 
$\hat{L}\hat{H}_2$ are gauge invariant
where $\hat{L}$ and $\hat{H}_{1,2}$ are the lepton doublet and Higgs 
doublet chiral superfields.
Thus we can construct two gauge invariant dimension three 
operators $\hat{S}_1\hat{H}_1\hat{H}_2$ and $\hat{N}\hat{L}\hat{H}_2$ 
by introducing singlet chiral superfields $\hat{S}_1$ and $\hat{N}$.
If we add $\hat{S}_1\hat{H}_1\hat{H}_2$ to the MSSM 
superpotential and remove
the ordinary $\mu$-term, we find that the global symmetry is 
extended.
There appear two new Abelian symmetries other than the $B$ and $L$ 
symmetry \cite{iq,dr,s}. 
They can be taken as the Peccei-Quinn symmetry and the $R$ symmetry. 
An example of their charge assignment for the fermionic components of 
the chiral superfields is shown 
in Table.\footnote{We
should note that $\hat{N}\hat{L}\hat{H}_2$ violates the conservation of 
$L$ for the charge assignment given in Table. 
This operator is important for both the explanation of the small
neutrino mass and the generation of the $B$ asymmetry. 
This point will be discussed later.}
Among these four Abelian symmetries, two global symmetries 
U(1)$_{B-L}$ and U(1)$_X$ remain as those with no 
SU(3) and SU(2) gauge anomaly.\footnote{It is easily checked that 
the U(1)$_X$ charge can be represented 
as the linear combination of the four global Abelian charges as 
$3X=B+L-10Q_{\rm PQ}+3Q_R$.} 
In this note we study a possibility that this $X$ charge asymmetry 
produced through the nonthermal process plays a role
as the origin of both abundances of the baryon and the dark matter.

\begin{figure}[tb]
\small
\begin{center}
\begin{tabular}{c|ccccccccccccc}
  &$\tilde g$, $\tilde W$, $\tilde B$  & $\hat{Q}_L$ &$\hat{\bar U}_L$ 
&$\hat{\bar D}_L$ & $\hat{L}_L$ 
 &$\hat{\bar E}_L$ &$\hat{H}_1$ &$\hat{H}_2$ & $\hat{S}_1$ &$\hat{N}$
 & $A_{SU(3)}$& $A_{SU(2)}$   \\ \hline
$Q_{\rm PQ}$ &  0 &  0 &$-2$ &1 & $-1$ &2 &$-1$ & 2 &  $-1$ &
  0 & $-{3\over 2}$ & $-1$\\
$Q_R$  &$-1$ & $-{1\over 6}$ &$-1$ & 0 &$-{5\over 6}$ 
&${2\over 3}$ &$-{5\over 6}$ & ${1\over 6}$ &$-{1\over 3}$ &
-${1\over 2}$ & $-5$ & $-{13\over 3}$\\
$B$  &  0 & ${1\over 3}$  &$-{1\over 3}$ & $-{1\over 3}$ & 
0 & 0 & 0 &0 & 0 & 0  &0 & ${3\over 2}$\\
$L$    & 0    & 0   & 0   & 0  & $1$   & $-1$ & 0   & 0 & 
0 & 0  & 0 & ${3\over 2}$   \\ \hline
\end{tabular}
\vspace*{3mm}\\
Table~ Global U(1) charge assignment for the fermionic components 
and gauge anomaly.
\end{center}
\end{figure}

\normalsize
We consider a model defined by a superpotential $W_{\rm MSSM}+W_1$.
$W_{\rm MSSM}$ is the superpotential for the MSSM Yukawa 
interactions
\begin{equation}
W_{\rm MSSM}=y_U^{\alpha\beta}\hat{Q}_\alpha\hat{\bar U}_\beta\hat{H}_2
+y_D^{\alpha\beta}\hat{Q}_\alpha\hat{\bar D}_\beta\hat{H}_1
+y_E^{\alpha\beta}\hat{L}_\alpha\hat{\bar E}_\beta\hat{H}_1.
\end{equation}
We introduce two massless MSSM singlet chiral superfields $\hat{S}_1$ and 
$\hat{S}_2$ for the construction of an additional superpotential 
$W_1$. It plays an essential role in the present scenario. 
We  assign the integer charge $q(>0)$ and 
$-p(<0)$ for the discrete symmetry $Z_m$ to $\hat{S}_1$ and $\hat{S}_2$, 
respectively. If the least common multiple for $p$ and $q$
is assumed to be $pq$ and $p+q=n~(n<m)$, a superpotential 
constructed from the lowest order $Z_m$ 
invariant operators can be written as
\begin{equation}
W_1=\lambda_1 \hat{S}_1\hat{H}_1\hat{H}_2 
+\lambda_2 \hat{S}_2\hat{F}_1\hat{F}_2
    +{d\over M_{\rm pl}^{n-3}}\hat{S}_1^p\hat{S}_2^q,    
\label{eqa}
\end{equation}
where couplings $\lambda_{1,2}$ and $d$ are assumed to be real and $d=O(1)$. 
We suppose that chiral superfields $\hat{H}_{1,2}$ and 
$\hat{F}_{1,2}$ have a suitable 
$Z_m$ charge so as to make the first two terms in $W_1$ be $Z_m$ invariant. 
$\hat{F}_{1,2}$ are assumed to have no MSSM gauge
interactions but have the global charge so that 
$\hat{S}_2\hat{F}_1\hat{F}_2$ is
invariant. They constitute a hidden sector which interacts with
the fields in the visible sector only through the gravity.
If we assign the chiral superfield $\hat S_2$ the global charges 
so as to be $X(\hat S_2)=x$
with $B=L=0$, SUSY breaking operators associated to the last term 
in $W_1$ may violate the $X$ conservation to give them 
the $X$ charge $\Delta X=4p+xq(\not=0)$. We will show that it can
generate the $X$ asymmetry through the AD mechanism \cite{ad}.
In the following part, the inflaton is assumed to couple directly with the
fields in the visible sector alone.

Now we consider a $D$-flat direction defined by 
$\langle S_1\rangle={u\over\sqrt q}e^{i\theta_1}$ and 
$\langle S_2\rangle ={u\over\sqrt p}e^{i\theta_2}$.\footnote{We note
that the potential minimum can be realized along the subspace 
$q|S_1|^2=p|S_2|^2$ as far as the soft SUSY breaking masses of $S_1$ and
$S_2$ are equal.}
This direction is slightly lifted by 
the nonrenormalizable operator in the scalar potential, 
which is induced from the last term of $W_1$. 
In the early universe, there are additional effective 
contributions to the scalar potential induced by the 
SUSY breaking effects caused by 
the large Hubble constant $H$ \cite{dens} and the thermal effects 
\cite{temp} other than the ordinary soft SUSY breakings.
If we take account of these effects, 
the scalar potential in that direction is found to be expressed as 
\begin{eqnarray}
V&\simeq&\left(-cH^2+M_u^2(T)\right)u^2
+{n\vert d\vert^2\over p^{q-1}q^{p-1}}
{u^{2n-2}\over M_{\rm pl}^{2n-6}} \nonumber\\ 
&+&{1\over p^{q\over 2}q^{p\over 2}}
\left\{\left({am_{3/2}e^{i\theta_a}\over M_{\rm pl}^{n-3}}
+{bHe^{i\theta_b} \over M_{\rm pl}^{n-3}}\right)
u^ne^{in\theta}+{\rm h.c.}\right\},
\label{eqaa}
\end{eqnarray}
where $\theta=(p\theta_1+q\theta_2)/n$. 
$m_{3/2}$ is a typical soft SUSY breaking scale of $O(1)$~TeV 
and the coefficients $a$, $b$ and $c$ are $O(1)$ real constants. 
CP phases $\theta_a$ and $\theta_b$ in the curly brackets 
are induced by the above mentioned SUSY breaking effects 
which violate the U(1)$_X$ by an amount of $\Delta X$.  
The effective mass $M_u^2(T)$ contains the usual soft SUSY 
breaking mass $m_{S}^2$ of $O(m_{3/2}^2)$ and the thermal mass 
$C_T\lambda_1^2T^2$ caused by the 
coupling of $\hat{S}_1$ with $\hat{H}_{1,2}$ 
in the thermal plasma.\footnote{Since the inflaton is assumed not to
couple directly with the hidden sector fields, no thermal plasma 
appears in the hidden sector.} 
It can be expressed depending on the value of $u$ as \cite{temp}
\begin{equation}
M_u^2(T)\simeq\left\{\begin{array}{ll}m^2_{S} & \quad (\lambda_1 u>T), \\
m^2_{S}+C_T\lambda_1^2 T^2 & \quad (\lambda_1 u<T), \\
\end{array}\right.
\label{therm}
\end{equation} 
where $C_T$ is a numerical factor in the thermal mass.

In the scalar potential (\ref{eqaa}), 
the Hubble constant contribution $H^2$ dominates the mass of the condensate
during the inflation.
If the sign of this Hubble constant contribution 
is negative $(c>0) $\cite{dens}, 
the magnitude of the condensate takes a large value such as
\begin{equation}
u_I\simeq \left(H M_{\rm pl}^{n-3}\right)^{1\over n-2}.
\label{eqa3}
\end{equation} 
On the other hand, the phase $\theta$ of the condensate at 
the potential minimum takes one of the $n$ distinct values 
$\theta=-\theta_b/n+2\pi \ell/n~(\ell=1, 2, \cdots, n)$.
At this period the condensate follows this instantaneous 
potential minimum since its evolution is almost the critical damping.
The dilute plasma appears as a result of a partial decay of the
inflaton. Then the temperature rapidly increases to 
$T_{\rm max}\simeq (T_R^2H M_{\rm pl})^{1/4}$ \cite{temp}.
$T_R$ is the reheating temperature realized after the completion 
of the inflaton decay and can be expressed as 
$T_R\simeq\sqrt{M_{\rm pl}\Gamma_I}$ where $\Gamma_I$ is the inflaton
decay width.
If this temperature $T_{\rm max}$ does not satisfy 
$\lambda_1\vert u_I\vert<T_{\rm max}$, 
no thermal contribution to $M_u^2(T)$ 
appears and $M_u^2(T)$ takes the expression of the upper one 
in eq.~(\ref{therm}) \cite{dens,temp}. 
Thus, the condition for the thermal effects to be negligible during the
inflation gives the following lower bound on $\lambda_1$:
\begin{equation}
\lambda_1 > T_R^{1/2}H_I^{n-6\over 4(n-2)}M_{\rm pl}^{10-3n\over 4(n-2)},
\label{eqa4}
\end{equation}
where $H_I$ is the Hubble parameter during the inflation.

When the inflaton evolves and $H$ decreases to $H\sim
m_{3/2}$,\footnote{Here we assume an inflation scenario 
in which this period is before the reheating. This means that 
$m_{3/2}>\Gamma_I$ is satisfied and then 
$T_R<\sqrt{M_{\rm pl}m_{3/2}}\simeq 10^{11}{\rm GeV}$.}  
the effective squared mass of the condensate becomes positive and 
then $u=0$ is the minimum of the scalar potential $V$.
The condensate starts to oscillate around $u=0$, and the
thermal effects due to the dilute plasma to $M_u^2(T)$ is expected to 
appear. 
Then $M_u^2(T)$ takes the expression of the lower one 
in eq.~(\ref{therm}). At this time, the dominant term for the 
U(1)$_X$ breaking changes from the second term in the curly brackets 
of eq.~(\ref{eqaa}) to the first one. 
Since the phase $\theta_a$ and $\theta_b$ are generally independent,
the phase $\theta$ of the condensate changes non-adiabatically from 
that determined by $\theta_b$ to that determined by $\theta_a$ 
due to the torque in the angular direction. 
Thus, the $X$ asymmetry is stored in the condensate during its 
evolution due to the AD mechanism \cite{ad}.

The produced $X$ asymmetry can be estimated by 
taking account that 
the $X$ current conservation is violated by the dominant $X$ breaking
operator in the curly brackets in eq.~(\ref{eqaa}) as
\begin{equation}
{d\Delta n_{X}(t)\over dt}= \Delta X
{am_{3/2}u^n\over M_{\rm pl}^{n-3}}\sin\delta,
\end{equation} 
where $\Delta X$ is the $X$ charge of that operator 
and $\delta$ is determined 
by the difference of $\theta_a$ and $\theta_b$. 
By solving this equation, the $X$ asymmetry produced in the condensates 
$\langle S_{1,2}\rangle$ at this period 
is found to be roughly expressed as 
\cite{ad,dens,temp}\footnote{The rigorous estimation requires the
numerical calculation as discussed in \cite{dens}. 
It is beyond the scope of this paper and we do not go further here.}
\begin{equation}
\Delta n_{X}(t)\simeq \Delta X{m_{3/2}\over H}H^{n\over n-2}
M_{\rm pl}^{2(n-3)\over n-2}\sin\delta,
\end{equation}
where $t$ is the time when $H\sim m_{3/2}$.  Following this, the
reheating due to the inflaton decay is completed at 
$H\sim \Gamma_I$.

The $X$ asymmetry stored in the condensate is 
liberated into the thermal plasma in the visible sector and also into the
hidden sector through the decay of the condensate 
by the $X$ conserving couplings $\lambda_1 \hat{S}_1\hat{H}_1\hat{H}_2$ and  
$\lambda_2 \hat{S}_2\hat{F}_1\hat{F}_2$, respectively.
We assume that the decay widths through these couplings satisfy
$\Gamma_{S_1}\simeq\Gamma_{S_2}$ to make the
discussion clear.\footnote{General cases will be discussed in Appendix.}
Since the oscillation of the condensate behaves as a matter for 
the expansion of the universe, it can dominate
the energy density of the universe before its decay which occurs at 
$H\sim \Gamma_{S_{1,2}}$.
For the reasonable value of $\lambda_{1,2}$, 
this is the case since $\Gamma_{S_{1,2}}<\Gamma_I$ is satisfied.  
Taking account of this, the ratio of the $X$ asymmetry $\Delta n_X^{V,H}$
liberated into each sector to the entropy density $s$ is estimated as
\begin{equation} 
Y_X^{V,H}\equiv{\Delta n_{X}^{V,H}(\tilde t_R)\over s}
={\Delta n_{X}(\tilde t_R)\over 2s}
\simeq {\Delta n_{X}(t)\over 2\tilde T_R^3}{t^2\over \tilde t_R^2}
\simeq {\Delta X\over 2}
\tilde T_Rm_{3/2}^{4-n\over n-2}M_{\rm pl}^{-2\over n-2}
\sin\delta, 
\label{eqe}
\end{equation}
where we use $\tilde t_R\sim 
\Gamma_{S_{1,2}}^{-1}\sim M_{\rm pl}/\tilde T_R^2$. 
If the temperature $\tilde T_R(\simeq 10^{10}\lambda_{1}{\rm GeV})$ 
is appropriate to keep the $X$ asymmetry\footnote{This 
$\tilde T_R$ is found to be a marginal value for the cosmological gravitino 
problem \cite{grav}.} and
convert it into the $B$ asymmetry through the sphaleron 
interaction \cite{spha}, we can obtain the $B$ asymmetry in the 
visible sector \cite{s}.
The $X$ asymmetry liberated in the hidden sector
is considered to explain the dark matter as far as the $X$ charge 
is conserved in the hidden sector.   

We examine whether the $X$ asymmetry in the visible sector 
transformed into the thermal plasma through the decay of the 
condensate can remain as a nonzero value and be partially 
converted into the $B$ asymmetry. 
This should be studied taking account that the electroweak 
sphaleron interaction and other various interactions are in 
the thermal equilibrium.
For this study, it is convenient to consider the detailed balance of
these interactions and solve
the chemical equilibrium equations \cite{lb,bp}. The particle-antiparticle
number asymmetry $\Delta n_f$ can be approximately related 
to the corresponding 
chemical potential $\mu_f$. In the case of $\mu_f \ll T$, it can be
represented as
\begin{equation}
\Delta n_f\equiv n_{f} -n_{f^c}=\left\{ \begin{array}{ll}
\displaystyle
{g_f\over 6}T^2\mu_f & (f~:~{\rm fermion}),\\
\displaystyle
{g_f\over 3}T^2\mu_f & (f~:~{\rm boson}), \\ \end{array} \right. 
\label{eqee}
\end{equation}
where $g_f$ is a number of relevant internal degrees of 
freedom of the field $f$.
By solving the detailed valance equations for the chemical potential $\mu_f$,
we can estimate the charge asymmetry at the period 
after the decay of the condensate.

If the SU(2) and SU(3) sphaleron interactions are in the thermal
equilibrium, we have the conditions such as
\begin{eqnarray}
&&\sum_{i=1}^{N_g}\left(3\mu_{Q_i}+\mu_{L_i}\right)+\mu_{\tilde{H_1}}
+\mu_{\tilde{H_2}}+ 4\mu_{\tilde W}=0,
\label{eqee1} \\  
&&\sum_{i=1}^{N_g}\left(2\mu_{Q_i}-\mu_{U_i}-\mu_{D_i}\right)
+6\mu_{\tilde g}=0,
\end{eqnarray}
where $N_g$ is a number of the generation of quarks and leptons.
The cancellation of the total hypercharge or the electric charge 
of plasma in the universe requires
\begin{eqnarray}
&&\sum_{i=1}^{N_g}\left(\mu_{Q_i}+2\mu_{U_i}-\mu_{D_i}-\mu_{L_i}-
\mu_{E_i}\right)+\mu_{\tilde{H_2}}-\mu_{\tilde{H_1}} \nonumber \\
&&\hspace{2cm}+2\sum_{i=1}^N(\mu_{\tilde{Q_i}}+2\mu_{\tilde{U_i}}-
\mu_{\tilde{D_i}}-\mu_{\tilde{L_i}}-\mu_{\tilde{E_i}})
+2\left(\mu_{H_2}-\mu_{H_1} \right)=0.
\end{eqnarray}
When Yukawa interactions in $W_{\rm MSSM}+W_1$ are in the thermal 
equilibrium, they impose the conditions\footnote{
We should note that the last term in $W_1$ leaves the thermal equilibrium 
at $T\sim M_{\rm pl}$. Since $\hat{S}_1$ has no other coupling to the MSSM
contents than $\lambda_1 \hat{S}_1\hat{H}_1\hat{H}_2$, 
the last one in eq.~(\ref{chemi}) is
the only condition for $\mu_{S_1}$.}
\begin{eqnarray} 
&&\mu_{Q_i}-\mu_{U_j}+\mu_{H_2}=0, \qquad
\mu_{Q_i}-\mu_{D_j}+\mu_{H_1}=0, \nonumber \\
&&\mu_{L_i}-\mu_{E_j}+\mu_{H_1}=0, \qquad
\mu_{S_1} +\mu_{\tilde H_1}+\mu_{\tilde H_2}=0.
\label{chemi}
\end{eqnarray}
There are also the conditions for the gauge interactions in the
thermal equilibrium, which are summarized as
\begin{equation}
\mu_{\tilde{Q_i}}=\mu_{\tilde g}+\mu_{Q_i}=\mu_{\tilde W}+\mu_{Q_i}
=\mu_{\tilde B}+\mu_{Q_i},
\label{eqee2}
\end{equation}
where $\mu_{\tilde g}$, $\mu_{\tilde W}$ and $\mu_{\tilde B}$ stand for
gauginos in the MSSM.
The similar relations to eq.~(\ref{eqee2}) is satisfied 
for leptons $\hat{L}_i$, Higgs fields $\hat{H}_{1,2}$ and other
fields $\hat{U}_i, \hat{D}_i,  \hat{E}_i $ which have the SM 
gauge interactions.
Flavor mixings of quarks and leptons due to the Yukawa couplings
allow us to consider the flavor independent chemical potential such as
$\mu_{Q}=\mu_{Q_i}$ and $\mu_{L}=\mu_{L_i}$.

Here we introduce an operator violating both $B-L$ and $X$, which is
necessary to convert the $X$ asymmetry into the $B$ and $L$ asymmetry.
If such an operator exists, only a linear combination of these two U(1)s is 
absolutely conserved. Then a part of $X$ asymmetry can be converted
into the $B-L$ asymmetry.
We consider an effective operator $(\hat{L}\hat{H}_2)^2$ as such an example.
It corresponds to the effective neutrino mass operator 
in the ordinary seesaw mechanism,
which is obtained from the operator $\hat{N}\hat{L}\hat{H}_2$ discussed
in the first part by integrating out the heavy right-handed neutrino 
$\hat{N}$.
The thermal equilibrium condition of this operator can be written as 
\begin{equation}
\mu_L+\mu_{H_2}=0.
\label{vlep}
\end{equation}

By now we have not taken account of the equilibrium 
conditions for soft SUSY breaking operators. 
The soft SUSY breaking operators are in the thermal equilibrium 
when $H~{^<_\sim}~\Gamma_{ss}$ is satisfied. 
Since the rate of the soft SUSY breaking effects is written as 
$\Gamma_{ss}\simeq m_{3/2}^2/T$ \cite{iq}, 
we find that the soft SUSY breaking operators are 
in the thermal equilibrium for the temperature 
$T~{^<_\sim}~T_{ss}\simeq 10^7$~GeV. 
Thus, for $T~{^<_\sim}~T_{ss}$ we find that $\mu_{\tilde g}=0$ is satisfied 
and then eqs.~(\ref{eqee1}) $\sim$ (\ref{vlep}) result 
in $\mu_{\tilde H_2}=0$. 
The $X$ asymmetry produced in the visible sector through the decay of
 the condensate disappears in this case. In order to escape this, 
if we define $T_X$ as a temperature
at which the $X$ and $B-L$ violating interaction is out-of-equilibrium,
we need to require that $T_X$ should satisfy 
$T_{ss}~{^<_\sim}~T_X~{^<_\sim}~\tilde T_R$.
In order that this condition is satisfied, 
the effective operator $(\hat{L}\hat{H}_2)^2$ have to leave the 
thermal equilibrium before the temperature reaches $T_{ss}$.
By using the right-handed neutrino mass $M_R$,
we can summarize this condition into a statement that 
$H>T_X^3/M^2_R$ should be satisfied at $T_X~{^>_\sim}~T_{ss}$. 
This results in $M_R~{^>_\sim}~10^{12}$~GeV, which is 
a suitable value for the explanation of the light neutrino masses 
required by the neutrino oscillation data \cite{sol,atm}.
In that case we find that there is an independent chemical potential in these 
thermal equilibrium conditions (\ref{eqee1})$\sim$(\ref{vlep}).
It can be taken as $\mu_{\tilde H_2}$, which corresponds to 
the above mentioned remaining symmetry.

The $X$ asymmetry induced in the
visible sector through the decay of the condensate can be partially 
converted into the $B$ asymmetry.
By solving eqs.~(\ref{eqee1}) $\sim$ (\ref{vlep}),
$\mu_Q$, $\mu_L$, $\mu_{H_{1,2}}$ and $\mu_{\tilde g}$ can be written 
with the chemical potential of Higgsino field $\tilde H_2$ 
at $T_X$ in such a way as
\begin{eqnarray}
&&\mu_Q={17N_g+6\over N_g(10N^2_g-17N_g-15)}\mu_{\tilde H_2}, \quad
\mu_L=-\mu_{H_2}={5(4N_g+3)\over 10N^2_g-17N_g-15}\mu_{\tilde H_2}, 
\nonumber \\
&&\mu_{H_1}=-{40N_g+3\over 10N^2_g-17N_g-15}\mu_{\tilde H_2}, \quad
\mu_{\tilde g}=-{(10N_g+3)N_g\over 10N^2_g-17N_g-15}\mu_{\tilde H_2}.
\label{eqee3}
\end{eqnarray}
Defining $B$ and $L$ as $\Delta n_B\equiv BT^2/6$ 
and $\Delta n_L\equiv LT^2/6$, 
we can calculate these values at $T_{ss}$ by using eqs.~(\ref{eqee}) and 
(\ref{eqee3}) as  
\begin{eqnarray}
&&B={80N_g^3+204N_g^2-150N_g-72\over 360N_g^3
+3308N^2_g-1419N_g-1143}X_1,  \nonumber \\  
&&L={N_g(60N^2_g-42N_g-126)\over 360N_g^3
+3308N^2_g-1419N_g-1143 }X_1, 
\label{eqeed}
\end{eqnarray}
where $X_1$ stands for the $X$ asymmetry stored in the $S_1$ condensate,
which is defined by $\Delta n^V_X\equiv X_1T^2/6$.
These results show that all of $B$, $L$ and $B-L$ take nonzero 
values as far as $X_1\not= 0$.

When the temperature goes below $T_{ss}$, 
the soft SUSY breaking operators are in the thermal equilibrium.
As mentioned above, this results in $\mu_{\tilde g}=0$ and $X_1=0$.
However, if the $X$ and $B-L$ violating interaction in the visible
sector is assumed to be out-of-equilibrium 
at $T_{ss}$, the equilibrium conditions are represented by
(\ref{eqee1})$\sim$(\ref{eqee2}). Thus the $B-L$ asymmetry existing at
$T_{ss}$ is kept after this period. The equilibrium conditions 
give the ordinary MSSM values for $B$ and $L$ as
\begin{equation}
B={4(2N_g+1)\over 22N_g+13}(B-L), \qquad L=-{14N_g+9\over 22N_g+13}(B-L),
\end{equation}
where we should use the $B-L$ value obtained from 
eq.~(\ref{eqeed}).
The $B$ asymmetry produced in this scenario is finally
estimated as
\begin{equation}
Y_B\equiv {\Delta n_B\over s} \simeq {\Delta n_X\over 2s}f(N_g)\kappa 
\simeq {\Delta X\over 2}\tilde T_R~m_{3/2}^{4-n\over n-2}
M_{\rm pl}^{-2\over n-2}f(N_g)~\kappa\sin\delta,
\label{eqe4}
\end{equation}
where eq.~(\ref{eqe}) is used and $\kappa (\le 1)$ is introduced 
to take account of the washout effect.
$f(N_g)$ is a numerical factor defined by
\begin{equation}
f(N_g)={B-L\over X_1}~{4(2N_g+1)\over 22N_g+13},
\end{equation}
and it takes $f(3)\simeq 0.3$ for $N_g=3$.
From this result, we find that this scenario can produce the presently
observed $B$ asymmetry $Y_B=(0.6~-~1)\times 10^{-10}$ as far as $n\ge 5$ and
in the case of $n=5$, for example, 
$\tilde T_R~{^>_\sim}~10^4/(\Delta X\kappa\sin\delta)$~GeV is required.

Since the $X$ asymmetry in the hidden sector is expected to explain the
dark matter, we can estimate the ratio of the baryon energy density to
the dark matter energy density by using eqs.~(\ref{eqe}) and (\ref{eqe4}) 
in such a way as
\begin{equation}
{\Omega_B\over \Omega_{\rm DM}}\simeq
{m_pY_B\over m_{\rm LP}Y_X^H}\simeq f(3)\kappa{m_p\over m_{\rm LP}},
\end{equation} 
where $\Omega_i$ is the ratio of the energy density $\rho_i$ to the
critical energy density $\rho_{\rm cr}$ in the universe.
Masses of the proton and the lightest stable particles in 
the hidden sector with nonzero $X$ charge are
represented by $m_p$ and $m_{\rm LP}$.
This relation suggests that the presently observed 
value $\Omega_B/\Omega_{\rm CDM}\sim 0.17$ can be explained 
if $m_{\rm LP}$ is the same order value as $m_p$ in the case of
$\kappa\simeq 1$.\footnote{This tuning of the mass of the 
dark matter field seems to be generally required and cannot be
avoided even in the case
that both number densities are related. An example evading this can be
found in \cite{qcdball} where the QCD-balls play the role 
of the dark matter.}
This value of $m_{\rm LP}$ also suggests that this dark
matter candidate behaves as the cold dark matter.  

Finally we order some remarks. 
Firstly, the evolution of $u$ may be able to induce the $\mu$-term \cite{s}.
We assume $H_I\sim 10^{13}$~GeV during the inflation on the basis of the CMB 
data and $T_R~{^<_\sim}~10^9$~GeV. For these values we obtain 
$T_{\rm max}\sim 10^{13}$~GeV. Thus, if we take $n=5$ in $W_1$, 
as an example, eq.~(\ref{eqa3}) gives $u_I\sim 10^{16}$~GeV and  
eq.~(\ref{eqa4}) suggests that $\lambda_1~{^>_\sim}~ 10^{-9}T_R^{1/2}$ should 
be satisfied. 
When the temperature decreases from $T_{\rm max}$ to 
$T_c\sim m_{3/2}/\lambda_1$, $M_{S_1}^2(T)$ represented by the
lower one in eq.~(\ref{therm}) starts to be dominated by the
 soft SUSY breaking mass $m_{S_1}^2$.
If $m_{S_1}^2<0$ is realized by some reason \cite{rad},
$u\not= 0$ becomes the true vacuum after this period. 
Since the $\mu$-term is generated from the first term in 
$W_1$ as $\mu=\lambda_1 u$,  
such a value of $u$ should be $u_0~{^<_\sim}~10^{11}T_R^{-1/2}$~GeV 
to realize the
appropriate $\mu$ for the above mentioned $\lambda_1$.\footnote{Such a
$u_0$ may be expected to be determined either by the nonrenormalizable 
terms or by the pure radiative symmetry breaking effect 
in which $u_0$ is estimated by using the renormalization group equation 
\cite{rad1}.}
Although the condensate again starts to oscillate around $u_0$, 
the deviation from $u_0$ instantaneously decays into the light fields 
through the $X$ conserving coupling $\hat{S}_1\hat{H}_1\hat{H}_2$ 
since $H<\Gamma_{S_1}$ is satisfied at this time.  
The released energy cannot dominate the total energy density
$({\pi^2\over 30}g_\ast T^4 \gg m_{3/2}^2u_0^2)$ and then the 
effects of the produced entropy is negligible.
Thus, even in this case the $X$ asymmetry obtained in eq.~(\ref{eqe}) 
can be used as the origin of the $B$ asymmetry. 

Secondly, it is useful to show a simple example of the mass generation
scenario to make the picture of the hidden sector a little bit clear, 
although there seem to be various possibilities of 
the structure for the hidden 
sector which can realize the right value for $m_{\rm LP}$ within the
present framework.\footnote{We will present 
another example in the extended version prepared in \cite{common}.} 
We consider the hidden sector composed of new chiral superfields 
$\hat C_{1,2}$ and $\hat S$ in addition to $\hat F_{1,2}$ and 
$\hat S_2$ in $W_1$.
If we suppose the $X$ charge assignment for these chiral superfields such as 
\begin{eqnarray}
&&X(\hat F_1)=1-x, \quad X(\hat F_2)=1, \quad 
X(\hat C_1)=1+x, \nonumber \\ 
&&X(\hat C_2)=2-x, \quad 
X(\hat S)=0, \quad X(\hat S_2)=x,
\end{eqnarray}
we can have the supplementary superpotential for these chiral 
superfields as
\begin{equation}
W_2=h_1{\hat S^2 \over M_{\rm pl}}\hat F_1\hat C_1
+h_2\hat S\hat S_2\hat C_2+{h_3\over 2}\hat S\hat F_2^2.
\end{equation}
We find that these operators are the lowest order ones 
allowed for the above $X$ charge assignment.\footnote{It should be noted that
the lower dimensional $X$ invariant renormalizable operators 
$\hat S_2\hat C_2$, $\hat F_2^2$ and $\hat S\hat F_1\hat C$ can be
forbidden by imposing the $Z_m$ symmetry appropriately.}
If $\langle S\rangle$ takes a nonzero value, fermions in the 
hidden sector get masses through the interactions in $W_2$ 
without breaking the $X$ charge. Their mass terms can be written as
\begin{equation}
\left(C_1, F_1\right)\left(\begin{array}{cc}
0 & h_1{\langle S\rangle^2 \over M_{\rm pl}}\\
h_1{\langle S\rangle^2 \over M_{\rm pl}} & 0\\
\end{array}\right)
\left(\begin{array}{c} C_1 \\ F_1 \\ \end{array} \right)
+\left(C_2, S_2\right)\left(\begin{array}{cc}
0 & h_2\langle S\rangle\\
h_2\langle S\rangle & 0\\\end{array}\right)
\left(\begin{array}{c} C_2 \\ S_2 \\ \end{array} \right)
+h_3\langle S\rangle F_2^2.
\end{equation}
where $\langle S_2\rangle$ is assumed to be zero since 
it breaks the $X$ charge. 
From this, we find that the lightest mass eigenvalue is
$m_{\rm LP}\simeq h_1\langle S\rangle^2/ M_{\rm pl}$.
This $m_{\rm LP}$ can be in a required region as far as 
$\langle S\rangle=O(10^{9})$~GeV is realized. 
Although the value of $\langle S\rangle$ depends on the more detailed
structure of the hidden sector, we can expect that it will occur in the
similar way discussed in \cite{rad1}.
The $X$ charge asymmetry distributed into the fermionic component of
$\hat F_1$ through the coupling $\lambda_2\hat S_2\hat F_1\hat F_2$
can present a suitable amount of the energy density of 
dark matter.\footnote{Although
the fermionic component of $\hat S$ is massless in the present example, 
$\hat F_1$ has no
renormalizable interaction with $\hat S$ and there is no effective decay mode
between them. Thus, the $X$ charge asymmetry
stored in the fermionic component of $\hat F_1$ is considered to be conserved.}

Thirdly, we remark the relation to the ordinary leptogenesis.
In the seesaw model the leptogenesis is usually considered 
on the basis of the out-of-equilibrium decay of the heavy right-handed 
neutrinos \cite{lept} or the decay of sneutrino condensate \cite{sneu}. 
However, if we consider the spontaneous $\mu$-term generation
along the almost flat direction of $\langle S_1\rangle$ as discussed above, 
the $B$ asymmetry produced by this usual leptogenesis 
might not be the dominant one. As mentioned before, we assume that 
the decay of the condensate is completed above the temperature 
$T_X$ which can be sufficiently lower
than the masses of the right-handed heavy neutrinos.
Then the $B$ asymmetry produced through the usual 
scenario seems to be washed out or overridden by the $B$ asymmetry 
produced in the present scenario.

Fourthly, we refer to the experimental signatures of the present model.
Since the dark matter field lives in the hidden sector in the present
model, it interacts with the fields in the observable sector 
only through the gravitational
interaction. This may make it discriminate experimentally
from other candidates in the MSSM. 
In the visible sector, we can also find the experimental signatures 
of this model in the the neutral Higgs and neutralino sectors.
Since this model is extended from the MSSM by the singlet chiral 
superfield $\hat S_1$ which has a coupling 
$\lambda_1\hat S_1\hat H_1\hat H_2$, the neutral
Higgs and neutralino sectors are changed from the MSSM. 
In the neutral Higgs scalar, there is an additional contribution for the
neutral Higgs mass from this coupling. Thus the lightest neutral Higgs
scalar can be heavier than that in the MSSM. In the neutralino sector also, 
some of the neutralinos can have an substantial ingredient coming 
from the fermionic component of $\hat S_1$. If we combine these 
features of the model, it may be possible to distinguish 
the present model from other proposals in the MSSM.
 
In summary, we have studied the possibility that both abundances of 
the baryon and the dark matter are originated from the asymmetry of the same
global charge in the supersymmetric framework.
In order to enlarge the global symmetry of the MSSM, we have introduced some
singlet chiral superfields. In that model we have shown that both
observed values of $Y_B$ and $\Omega_B/\Omega_{\rm CDM}$ could be
realized through the global charge asymmetry stored in the condensate of
these singlet scalar components. 
This kind of possibility for the production of the $B$ asymmetry and the
dark matter may be worth to further study as much as the usual scenario.

\vspace*{5mm}
\noindent
This work is supported in part by a Grant-in-Aid for Scientific 
Research (C) from Japan Society for Promotion of Science (No.~14540251).

\vspace*{1cm}
\noindent
{\Large\bf Appendix}

\noindent
In this appendix we discuss the amount of $X$ asymmetry liberated into
each sector in general cases with arbitrary $\Gamma_{S_1}$ and 
$\Gamma_{S_2}$.
If we take account that the decay products of the condensate behave as
the radiation and also no photon is produced through its decay into the 
hidden sector, we can estimate the $X$ asymmetry 
$\Delta n_X^V(\tilde t_R)$ generated in the visible sector in case
$\Gamma_{S_1}>\Gamma_{S_2}$ as,
\begin{eqnarray} 
{\Delta n_{X}^V(\tilde t_R)\over s}&=&
{\Delta n_{X}(t)\over s}
{\Gamma_{S_1}\over \Gamma_{S_1}+\Gamma_{S_2}}
\left({t\over t_1}\right)^2,
\end{eqnarray}
and also in case $\Gamma_{S_1}<\Gamma_{S_2}$ as,
\begin{eqnarray} 
{\Delta n_{X}^V(\tilde t_R)\over s}&=&
{\Delta n_{X}(t)\over s}
{\Gamma_{S_1}\over \Gamma_{S_1}+\Gamma_{S_2}}
\left({t\over t_2}\right)^2\left({t_2\over t_1}\right)^{3/2},
\end{eqnarray}
where $\tilde t_R\simeq t_1=\Gamma_{S_1}^{-1}$ and $t_2= \Gamma_{S_2}^{-1}$.
Since we know from the discussion in the text that 
the $B$ asymmetry $Y_B$ can be expressed by using 
$\Delta n_X^V(\tilde t_R)$ as
\begin{equation}
Y_B\equiv {\Delta n_B\over s} \simeq {\Delta n_X^V(\tilde t_R)\over s}
~f(N_g)~\kappa, 
\end{equation}
we can obtain $Y_B$ in each case as follows:
\begin{eqnarray}
&&Y_B\simeq \Delta X~ \tilde T_R~m_{3/2}^{4-n\over n-2}
M_{\rm pl}^{-2\over n-2}f(N_g)~\kappa\sin\delta 
\qquad (\Gamma_{S_1}>\Gamma_{S_2}), \nonumber \\
&&Y_B\simeq \Delta X~ \tilde T_R~m_{3/2}^{4-n\over n-2}
M_{\rm pl}^{-2\over n-2}\left({\Gamma_{S_1}\over\Gamma_{S_2}}\right)^{1/2}
f(N_g)~\kappa\sin\delta
\qquad (\Gamma_{S_1}<\Gamma_{S_2}).
\end{eqnarray}
Although  $Y_B$ has the same expression as that discussed in the text
for the case of $\Gamma_{S_1}>\Gamma_{S_2}$, 
there is an additional suppression factor 
$(\Gamma_{S_1}/\Gamma_{S_2})^{1/2}$ for the case of 
$\Gamma_{S_1}<\Gamma_{S_2}$. 
Thus, if we impose this case to realize the observed $B$ asymmetry for
the fixed values of $n$, $\kappa$ and $\sin\delta$,
the higher reheating temperature $\tilde T_R$ is required in comparison 
with the case of $\Gamma_{S_1}~{^>_\sim}~\Gamma_{S_2}$.

On the other hand, the ratio of the energy density of the baryon and the
dark matter can be expressed in both cases as
\begin{equation}
{\Omega_B\over \Omega_{\rm DM}}\simeq
{m_pY_B\over m_{\rm LP}Y_{X}^H}\simeq f(3)\kappa
{m_p\over m_{\rm LP}}{\Gamma_{S_1}\over\Gamma_{S_2}}.
\end{equation} 
This suggests that the presently observed value of
$\Omega_B/\Omega_{\rm DM}$ seems to be explained as far as 
$m_{\rm LP}\simeq (\Gamma_{S_1}/\Gamma_{S_2})\kappa$~GeV is satisfied. 
As far as we set up the hidden sector suitably, these lightest stable
particles in the hidden sector are expected to behave as the required 
cold dark matter.

\newpage

\end{document}